\documentclass{ws-mpla}
\usepackage{graphicx}
\usepackage{color}
\begin{document}

\markboth{E F D Evangelista \& J C N de Araujo}
{Stochastic Background of Gravitational Waves}

\catchline{}{}{}{}{}

\title{A NEW METHOD TO CALCULATE THE STOCHASTIC BACKGROUND OF GRAVITATIONAL WAVES GENERATED BY COMPACT BINARIES}

\author{\footnotesize EDGARD F D EVANGELISTA}

\address{Instituto Nacional de Pesquisas Espaciais -- Divis\~{a}o de Astrof\'{i}sica,\\ Av. dos Astronautas 1758, S\~{a}o Jos\'{e} dos Campos, 12227-010 SP, Brazil \\
edgard.evangelista@inpe.br}

\author{\footnotesize JOS\'{E} C N DE ARAUJO}

\address{Instituto Nacional de Pesquisas Espaciais -- Divis\~{a}o de Astrof\'{i}sica,\\ Av. dos Astronautas 1758, S\~{a}o Jos\'{e} dos Campos, 12227-010 SP, Brazil \\
jcarlos.dearaujo@inpe.br}

\maketitle

\pub{Received (Day Month Year)}{Revised (Day Month Year)}

\begin{abstract}
In the study of gravitational waves (GWs), the stochastic background generated by compact binary systems are among the most important kinds of signals. The reason for such an importance has to do with their probable detection by the interferometric detectors [such as the Advanced LIGO (ALIGO) and Einstein Telescope (ET)] in the near future. In this paper we are concerned with, in particular, the stochastic background of GWs generated by double neutron star (DNS) systems in circular orbits during their periodic and quasi--periodic phases. Our aim here is to describe a new method to calculate such spectra, which is based on an analogy with a problem of Statistical Mechanics. Besides, an important characteristic of our method is to consider the time evolution of the orbital parameters.

\keywords{Gravitational waves; stochastic background; double neutron stars.}
\end{abstract}

\ccode{PACS Nos.: 04.30.Db, 95.85.Sz}

\section{Introduction}

A big challenge of modern Astrophysics is the detection of gravitational waves (GWs), which would provide a new window to observe the universe. In the same way we use electromagnetic radiation such as X-rays, $\gamma$-rays, infrared or visible light to study the astrophysical objects, we could well use the observation of gravitational radiation as an efficient tool in the search of the properties of the several kinds of astronomical objects. Specifically, we are concerned with the stochastic background of GWs generated by double neutron star (DNS) systems. The study of this kind of signal is important because it is among the most probable sources to be detected in the near future or, putting in other words, these signals could form a foreground for the planned GW interferometers eLISA, BBO, DECIGO, Einstein Telescope (ET) and Advanced LIGO (ALIGO).

Therefore, given the importance of the stochastic background generated by DNS, we use it
as an example of application of a new method, described in the present paper, to calculate such spectra. The starting point is the equation\cite{araujo05}

\begin{equation}
h_{\mbox{\scriptsize{BG}}}^{2}=\frac{1}{\nu_{\mbox{\scriptsize{obs}}}}\int h_{\mbox{\scriptsize{source}}}^{2}dR
\label{bg}
\end{equation}
where $h_{\mbox{\scriptsize{BG}}}$ represents the dimensionless amplitude of the spectrum, $\nu_{\mbox{\scriptsize{obs}}}$ is the observed frequency, $dR$ is the differential rate of generation of gravitational radiation and $h_{\mbox{\scriptsize{source}}}$ is the amplitude of the emitted radiation by a single source, namely\cite{evans,thorne}

\begin{equation}
h_{\mbox{\scriptsize{source}}}=7.6\times 10^{-23}\left(\frac{\mu}{M_{\odot}}\right)\left(\frac{M}{M_{\odot}}\right)^{2/3}\left(\frac{1\mbox{Mpc}}{d_{\mbox{\scriptsize{L}}}}\right)\left(\frac{\nu}{1\mbox{Hz}}\right)^{2/3}
\label{source}
\end{equation}
where $\mu$ is the reduced mass of the system, $M$ is the total mass and $d_{\mbox{\scriptsize{L}}}$ is the luminosity distance.

It is worth mentioning that Eq.~(\ref{bg}) was obtained from an energy-flux equation. This equation was first derived in a paper by de Araujo et al\cite{araujo00} and was used in their various papers. In particular, in a subsequent paper\cite{araujo05} they gave a more detailed derivation of this equation, showing its robustness. Also, although apparently simple it contains the correct and necessary ingredients to calculate the background of GWs for a given type of source.

Given Eq.~(\ref{bg}) and once $h_{\mbox{\scriptsize{source}}}$ is known, we are focused on the calculation of the rate $dR$. First, we write this rate in the form

\begin{equation}
dR=\frac{dR}{dV}\frac{dV}{dz}dz
\label{dr}
\end{equation}
where $dV$ is the comoving volume element and $z$ is the redshift. The element $dV/dz$ is known from cosmology (see Subsection \ref{cosmology}) and $dR/dV$ will be calculated by means of the new method considered in this paper.

So, in order to describe the process to obtain the stochastic background, we organized this paper as follows: in Section \ref{sec2} we show the population characteristics of DNS and the elements of cosmology necessary to the calculation of the spectra; in Section \ref{sec3} we explain the method itself; the results are shown and discussed in Section \ref{sec4}; and in Section \ref{sec5} we present the conclusions and perspectives.

\section{Population Characteristics of DNS}
\label{sec2}

For the initial mass function (IMF), which is one of the ingredients of our calculations, we adopt the Salpeter distribution, namely\cite{salpeter}

\begin{equation}
\phi(m)=Am^{-(1+x)}
\label{salpeter1}
\end{equation}
where $A=0.17$ is the normalization constant and $x=1.7$. The normalization of this IMF is obtained by means of

\begin{equation}
\int^{m_{f}}_{m_{i}}m\phi(m)dm=1
\end{equation}
where we are considering $m_{f}=125\mbox{M}_{\odot}$ and $m_{i}=0.1\mbox{M}_{\odot}$\cite{ostlie}. Furthermore, we consider that neutron stars are generated by progenitors with masses ranging from $8\mbox{M}_{\odot}$ to $25\mbox{M}_{\odot}$. It is worth pointing out that, although the choice of the values for these minimum and maximum masses of the progenitors are subject of discussion, the range $8-25\mbox{M}_{\odot}$ is usual in the literature (see, for example, Ferrari et al.\cite{ferrari1,ferrari2}, where the authors used these values). Further, discussions on this issue can also be found, for example, in Smartt\cite{smartt} and Carrol \& Ostlie\cite{ostlie}.

Besides the IMF, we need to adopt a star formation rate density (SFRD), which is another ingredient that appears in the calculation of the background. There are, in the literature, many alternatives for such ingredient. In particular, we adopt the SFRD derived by Springel \& Hernquist\cite{springel}, namely

\begin{equation}
\dot{\rho}_{\ast}(z)=\dot{\rho}_{m}\frac{\beta e^{\alpha(z-z_{m})}}{\beta-\alpha+\alpha e^{\beta(z-z_{m})}}
\label{starfor}
\end{equation}
where $\alpha=3/5$, $\beta=14/15$, $z_{m}=5.4$ and with $\rho_{m}=0.15\mbox{M}_{\odot}\mbox{yr}^{-1}\mbox{Mpc}^{-3}$ fixing the normalization. It is worth mentioning that Eq.~(\ref{starfor}) was obtained considering  a $\Lambda$CDM cosmology in a structure formation scenario, where the density parameters have the values $\Omega_{\mbox{\scriptsize{M}}}=0.3$, $\Omega_{\mbox{\scriptsize{B}}}=0.04$ and $\Omega_{\Lambda}=0.7$; where the subscripts $M$, $B$ and $\Lambda$ referes to matter, baryonic matter and cosmological constant, respectively. Besides, the authors used for Hubble's constant the value of $H_{0}=100h \,\mbox{km}\,\mbox{s}^{-1}\mbox{Mpc}^{-1}$ with $h=0.7$ and it was considered a scale invariant power spectrum with index $n=1$, normalized to the abundance of rich galaxy clusters at present day ($\sigma_{8}=0.9$).

It is worth mentioning that one could argue that a different choice for the SFRD would modify significantly our results and conclusions. Since we are here mainly concerned with a new method to calculate the stochastic background of GWs, we leave the discussion concerning how different SFRDs affect the spectrum of the background of GWs, among other issues, to another paper to appear elsewhere.

With the IMF and the SFRD at hand, it is possible to determine the formation rate of DNS. First, we consider the mass fraction $\lambda_{\mbox{\scriptsize{nsns}}}$ that is converted into double neutron star systems\cite{regimbau2}, namely

\begin{equation}
    \lambda_{\mbox{\scriptsize{nsns}}}=\beta_{\mbox{\scriptsize{ns}}}f_{p}\Phi_{\mbox{\scriptsize{ns}}}
\end{equation}
where $\beta_{\mbox{\scriptsize{ns}}}$ is the fraction of binary systems that survive to the second supernova event, $f_{p}$ gives us the fraction of massive binaries (that is, binary systems where both components could generate a supernova event) formed from the whole population of stars and $\Phi_{\mbox{\scriptsize{ns}}}$ is the mass fraction of neutron star progenitors that, in our case and using Eq.~(\ref{salpeter1}), is given by

\begin{equation}
    \Phi_{\mbox{\scriptsize{ns}}}=\int^{25}_{8}\phi(m)dm.
\end{equation}

Numerically, one obtains  $\Phi_{\mbox{\scriptsize{ns}}}=5.97\times 10^{-3}\mbox{M}_{\odot}^{-1}$.
Following a paper by Regimbau and de Freitas Pacheco\cite{regimbau2}, one has $\beta_{\mbox{\scriptsize{ns}}}=0.024$
and $f_{p}=0.136$.
Using these results, the binary formation rate for DNS is given by

\begin{equation}
n_{\mbox{\scriptsize{nsns}}}(z)=\lambda_{\mbox{\scriptsize{nsns}}}\dot{\rho}_{\ast}(z)\mbox{.}
\label{bin_form}
\end{equation}

Concerning the orbital parameters of the DNS, for the sake of simplicity, we consider a uniform period distribution given by\cite{hils}

\begin{equation}
f(P)=\left\{\begin{array}{rc}
&(P_{\mbox{\scriptsize{u}}}-P_{\mbox{\scriptsize{l}}})^{-1} \hspace{14pt}P_{\mbox{\scriptsize{l}}}\leq P\leq P_{\mbox{\scriptsize{u}}}\\
&0 \hspace{47pt}\mbox{otherwise}
\end{array}\right.
\label{distr1}
\end{equation}
where $P_{\mbox{\scriptsize{u}}}$ and $P_{\mbox{\scriptsize{l}}}$ are the maximum and minimum periods, respectively.

Since our main aim here is to present an alternative method to calculate stochastic background, the choice above is suitable for this purpose.

However, in order to be used in the calculation of the spectra, Eq.~(\ref{distr1}) should be rewritten in terms of the frequency. This is achieved by changing variables via $f(P)dP=g(\nu_{\mbox{\scriptsize{orb}}})d\nu_{\mbox{\scriptsize{orb}}}$, where the period $P$ and the orbital frequency $\nu_{\mbox{\scriptsize{orb}}}$ are related to each other by $P=\nu_{\mbox{\scriptsize{orb}}}^{-1}$. Algebraic manipulations yield

\begin{equation}
g(\nu_{\mbox{\scriptsize{orb}}})=\left(\frac{1}{P_{\mbox{\scriptsize{u}}}-P_{\mbox{\scriptsize{l}}}}\right)\frac{1}{\nu_{\mbox{\scriptsize{orb}}}^{2}} \hspace{14pt}\mbox{for}\hspace{14pt}P_{\mbox{\scriptsize{l}}}\leq P\leq P_{\mbox{\scriptsize{u}}}
\label{distr2}
\end{equation}

Now, note that the frequency $\nu_{\mbox{\scriptsize{orb}}}$ undergoes time evolution. Following the paper by Peters\cite{peters} one obtains

\begin{equation}
\nu_{\mbox{\scriptsize{orb}}}=\frac{1}{2\pi}\left[(2\pi\nu_{\mbox{\scriptsize{orb}},0})^{-8/3}-\frac{8}{3}K(t-t_{0})\right]^{-3/8}
\label{freqev}
\end{equation}
where $K=96m_{1}m_{2}5c^{5}G^{5/3}(m_{1}+m_{2})^{-1/3}$, $m_{1}$ and $m_{2}$ are the masses of the components of the system and $\nu_{\mbox{\scriptsize{orb}},0}$ is the initial frequency. Therefore, we should carry out a further change of variables in Eq.~(\ref{distr2}) in order to include the time dependence given by Eq.~(\ref{freqev}). This is obtained by means of $g(\nu_{\mbox{\scriptsize{orb}},0})d\nu_{\mbox{\scriptsize{orb}},0}=H(\nu_{\mbox{\scriptsize{orb}}})d\nu_{\mbox{\scriptsize{orb}}}$ where $\nu_{\mbox{\scriptsize{orb}},0}$ is the initial frequency, which was associated with the variable $\nu_{\mbox{\scriptsize{orb}}}$ in (\ref{distr2}) and $H(\nu_{\mbox{\scriptsize{orb}}})$ is the new distribution which, after some algebra, is written as

\begin{equation}
H(\nu_{\mbox{\scriptsize{orb}}})=\left(\frac{1}{P_{\mbox{\scriptsize{u}}}-P_{\mbox{\scriptsize{l}}}}\right)(\nu_{\mbox{\scriptsize{orb}},0})^{5/3}(\nu_{\mbox{\scriptsize{orb}}})^{-11/3}
\label{dist1}
\end{equation}

Notice that it would be necessary to perform a further coordinate transformation in $H(\nu_{\mbox{\scriptsize{orb}}})$ in order to put it as a function of the emitted frequency $\nu$ instead of $\nu_{\mbox{\scriptsize{orb}}}$. Such a transformation, given by $\nu=2\nu_{\mbox{\scriptsize{orb}}}$, is trivial and all the equations will be written as functions of $\nu$ from now on. Still concerning $\nu$, it is related to the observed frequency $\nu_{\mbox{\scriptsize{obs}}}$ by means of $\nu=\nu_{\mbox{\scriptsize{obs}}}(1+z)$.

\subsection{Cosmology}
\label{cosmology}

To perform the calculation of the background of GWs it is necessary to specify the cosmology
and its corresponding parameters. Here we consider a flat universe.

An essential quantity is the comoving volume element, which is given by

\begin{equation}
\frac{dV}{dz}=4\pi \left(\frac{c}{H_{0}}\right)r_{z}^{2}F(\Omega_{\mbox{\scriptsize{M}}},\Omega_{\mbox{\scriptsize{$\Lambda$}}},z),
\end{equation}
where

\begin{equation}
F(\Omega_{\mbox{\scriptsize{M}}},\Omega_{\Lambda},z)=\frac{1}{\sqrt{(1+z)^{2}(1+\Omega_{\mbox{\scriptsize{M}}}z)-z(2+z)\Omega_{\Lambda}}},
\end{equation}
and the comoving distance $r_{z}$ for a flat universe reads

\begin{equation}
r_{z}=\frac{c}{H_{0}}
\int_{0}^{z}F(\Omega_{\mbox{\scriptsize{M}}},\Omega_{\Lambda},z')dz'
\end{equation}
where the density parameters obey the following relations:

\begin{equation}
\Omega_{\mbox{\scriptsize{M}}}=\Omega_{\mbox{\scriptsize{DM}}}+\Omega_{\mbox{\scriptsize{B}}}\:\:\:\mbox{and}\:\:\:\Omega_{\mbox{\scriptsize{M}}}+\Omega_{\Lambda}=1
\end{equation}
The values of these parameters are those presented in the previous section.

\section{The method}
\label{sec3}

Since we are considering the time evolution of the orbital frequency of the systems, we need
to take this issue into account in the derivation of the rate $dR/dV$ in Eq.~(\ref{dr}). We will see that the derivation of this rate comes down to count systems that reach a given
frequency at a given moment of time.

\begin{figure}[!htbp]
\centerline{\includegraphics[width=3.0in]{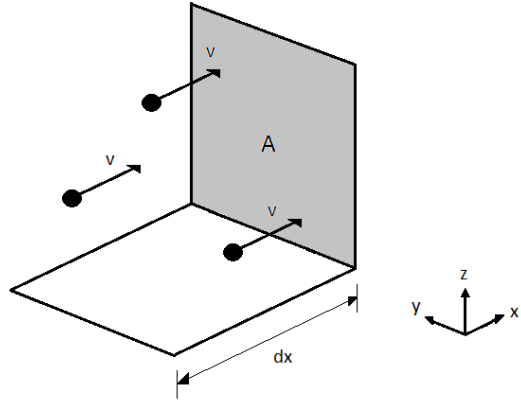}}
\vspace*{8pt}
\caption{The volume element $dV$ adjacent to $A$.\label{VolEl}}
\end{figure}

We derive this rate by means of an analogy with a problem of Statistical Mechanics. In this problem, the aim is to calculate the number of particles that reach a given area $A$ in a time interval $dt$, that is, the objective is to calculate the flux $F$ of particles. Basically, this flux is calculated by summing all the particles inside the volume $dV=Adx$ adjacent to the area $A$ and that move towards $A$ with velocity $v=dx/dt$, where $v$ obeys a distribution function $\eta(v)$ (see Fig.~\ref{VolEl}). Hence, the sum is obtained by integrating over all the positive values of $v$.

With some modifications, the method to calculate the flux $F$ can be used to determine $dR/dV$. First, we substituted the spatial coordinate $x$ by the frequency $\nu$ and the velocity $v$ by the time variation of the frequency, which is defined by $\upsilon_{\nu}=d\nu/dt$. So, the number of systems in the interval $d\nu$ adjacent to a particular frequency $\nu$ is given by

\begin{equation}
    \psi(\nu)=\frac{\varphi(\nu)}{\int \varphi(\nu^{\prime})d\nu^{\prime}}
\label{back2}
\end{equation}
where $\varphi(\nu)$ is the non-normalized distribution of frequencies.

Considering that the distribution $\eta(\upsilon_{\nu})$ gives the number of systems which have $\upsilon_{\nu}$ in the interval $d\upsilon_{\nu}$, the number of systems in $d\nu$ and with values of $\upsilon_{\nu}$ in the interval $d\upsilon_{\nu}$ is given by

\begin{equation}
    d\mu=\left(\frac{\varphi(\nu)d\nu}{\int \varphi(\nu^{\prime})d\nu^{\prime}}\right)\eta(\upsilon_{\nu})d\upsilon_{\nu}.
\label{dmu}
\end{equation}
Now, the next step is to determine the forms of $\varphi(\nu)$ and $\eta(\upsilon_{\nu})$. First, the distribution $\varphi(\nu)$ is written in the form

\begin{equation}
\label{inte}
   \varphi(\nu) =\int n_{\mbox{\scriptsize{nsns}}}(t_{o})H(\nu)dt_{o}
\end{equation}
where $t_{0}$ is the instant of birth  of the systems, $n_{\mbox{\scriptsize{nsns}}}$ is the formation rate density of the DNS and $H(\nu)$ is given by Eq.~(\ref{dist1}).

In the deduction of Eq.~(\ref{inte}) we consider initially $H(\nu)$, from which we have

\begin{equation}
dn=H(\nu)d\nu ,
\end{equation}
which is the fraction of systems originated at the time $t_{0}$ and that have frequencies in the interval $d\nu$. Now, using $n_{\mbox{\scriptsize{nsns}}}$, we can write explicitly

\begin{equation}
\frac{dn}{d\nu dVdt_{0}}=n_{\mbox{\scriptsize{nsns}}}(t_{0})H(\nu).
\end{equation}
Now, integrating over $dt_{0}$, we get

\begin{equation}
\frac{dn}{dV}=\left[\int^{t}_{t_{\mbox{\tiny{min}}}} n_{\mbox{\scriptsize{nsns}}}(t_{0})H(\nu)dt_{0}\right]d\nu ,
\label{app1}
\end{equation}
where the expression in brackets is the number of systems per unit frequency interval and per comoving volume at given time (or redshift), which is the desired distribution function $\varphi(\nu)$.

In Eq.~(\ref{app1}), the limits $t$ and $t_{\mbox{\scriptsize{min}}}$ of the integral are related to the redshifts $z$ and $z_{\mbox{\scriptsize{sup}}}$, respectively, with the usual expression found in any textbook on cosmology; further, $z_{\mbox{\scriptsize{sup}}}$ (see, e.g., Regimbau and Mandic 2008\cite{regman}) is given by
\begin{equation}
\label{eq13}
z_{\mbox{\scriptsize{sup}}}=\left\{\begin{array}{l}
z_{\mbox{\scriptsize{max}}} \hspace{20pt}\mbox{if} \hspace{20pt}\nu_{\mbox{\scriptsize{obs}}}<\dfrac{\nu_{\mbox{\scriptsize{max}}}}{1+z_{\mbox{\scriptsize{max}}}}\\
\vspace{2pt}\\
\dfrac{\nu_{\mbox{\scriptsize{max}}}}{\nu_{\mbox{\scriptsize{obs}}}}-1 \hspace{15pt}\mbox{otherwise.}
\end{array}\right.
\end{equation}
It is worth mentioning that we are considering $\nu_{\mbox{\scriptsize{max}}}=900$Hz (see next Section) and $z_{\mbox{\scriptsize{max}}}=20$.

On the other hand, $\eta(\upsilon_{\nu})$ will have a peculiar form. First, note that the derivation of Eq.~(\ref{freqev}) yields

\begin{equation}
\upsilon_{\nu}\equiv \frac{d\nu}{dt}\propto \nu^{\frac{11}{3}}
\end{equation}
after some algebraic manipulations. Then, we conclude that there will be just one value of $\upsilon_{\nu}$ for each value of $\nu$, which allows us to write $\eta(\upsilon_{\nu})$ as a Dirac delta function:
\begin{equation}
\label{xi}
    \eta(\upsilon_{\nu})=N\delta(\upsilon_{\nu}-\upsilon_{\nu,p})
\end{equation}
where $N$ is the total number of systems and $\upsilon_{\nu,p}$ is the particular value of $\upsilon_{\nu}$ corresponding to each frequency $\nu$.

Now, noting that the denominator of the term between parenthesis in Eq.~(\ref{back2}) is the total number of systems, using the function given by Eq.~(\ref{xi}) and changing the differential $d\nu$ by means of the chain rule, Eq.~(\ref{dmu}) assumes the form

\begin{equation}
d\mu=\left(\frac{\varphi(\nu)\frac{d\nu}{dt}dt}{N}\right)N\delta(\upsilon_{\nu}-\upsilon_{\nu,p})d\upsilon_{\nu}.
\end{equation}
Integrating over $\upsilon_{\nu}$ and rearranging the result, we obtain

\begin{equation}
R=\varphi(\nu)\frac{d\nu}{dt}
\end{equation}
where $R$ is the number of systems per time interval $dt$. Recalling that the rate $R$ is per comoving volume, we may write

\begin{equation}
\label{bg4}
\varphi(\nu)\frac{d\nu}{dt}\equiv \frac{dR}{dV}
\end{equation}
and Eq.~(\ref{bg}) assumes the form

\begin{equation}
\label{bg2}
h_{\mbox{\scriptsize{BG}}}^{2}=\frac{1}{\nu_{\mbox{\scriptsize{obs}}}}\int h_{\mbox{\scriptsize{source}}}^{2}\frac{dR}{dV} \frac{dV}{1+z} ,
\end{equation}
where we included the term $(1+z)$ in order to consider the time dilation due to the expansion of the universe.
Further, using this amplitude we can obtain the spectral amplitude, which is given by:
\begin{equation}
S_{h}=\frac{h_{\mbox{\scriptsize{BG}}}^{2}}{\nu_{\mbox{\scriptsize{obs}}}}.
\end{equation}

\section{Results}
\label{sec4}

Recall that the main aim here is to see how the time evolution of the orbital frequency affect the spectra of the GW background. Then we considered two cases: first, we used $10^{-6}$ and $10^{-4}$Hz as the minimum and the maximum frequencies (corresponding to periods of $P_{\mbox{\scriptsize{u}}}=10^{7}$sec and $P_{\mbox{\scriptsize{l}}}=10^{5}$sec, respectively); in the second case we considered $10^{-5}$ and $10^{-3}$Hz (which give us $P_{\mbox{\scriptsize{u}}}=10^{6}$sec and $P_{\mbox{\scriptsize{l}}}=10^{4}$sec). Note that here $P_{\mbox{\scriptsize{u}}}$ and $P_{\mbox{\scriptsize{l}}}$ refer to the period of the waves, while in Eq.~{\ref{dist1}} such parameters refer to the orbital period. The transformation is trivial in this case.

\begin{figure}[!htbp]
\centerline{\includegraphics[width=3.0in,angle=270]{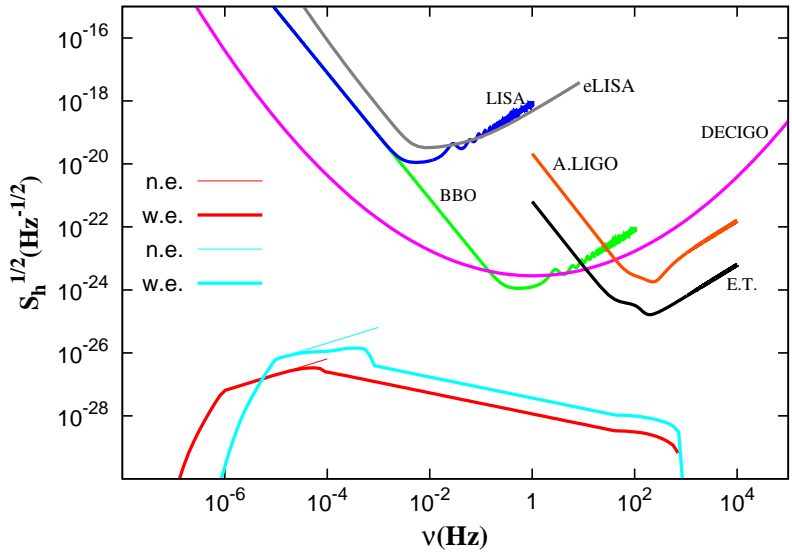}}
\vspace*{8pt}
\caption{Spectra generated with the use of Eq.~(\ref{dist1}). Note that in each case, in the regions of lower frequencies, there is superposition between the curves generated with (w.e.) and without evolution (n.e.). Also shown for comparison the sensitivity curves for LISA, eLISA, BBO, DECIGO, ET and ALIGO. \label{espectro1}}
\end{figure}

Figure (\ref{espectro1}) shows the spectra we generated using the distribution given by Eq.~(\ref{dist1}). Further, in order to point out the effects of the time evolution of the frequency on the results, we plotted in the same figure the curves corresponding to the case where the evolution is not considered.

In both cases, we can observe the influence of the evolution of the frequencies on the form of the spectra. Particularly, if the evolution is taken into account, there will be a reduction in the amplitude of the spectra at the region of maximum initial frequency and a spread towards higher frequencies. Concerning the maximum frequency of the radiation emitted by the systems, we are assuming the value $\nu_{\mbox{\scriptsize{max}}}=900$Hz\cite{isco}.

Fig.~(\ref{espectro1}) also shows for comparison the sensitivity curves for LISA, eLISA\cite{elisa}, BBO\cite{bbo}, DECIGO\cite{decigo}, ET\cite{satya} and ALIGO\cite{satya}. The sensitivity curve for LISA may be found at http://www.srl.caltech.edu/\verb|~|shane/sensitivity/.

Although the spectra shown in Fig.~(\ref{espectro1}) do not form foregrounds and therefore cannot be detected by single interferometric detectors, a suitable correlation of two or more of such detectors could, in principle, detect this background.\cite{michelson,romano,allen} In fact, the analysis of the detectability by cross-correlation, among other issues, are discussed in our other paper\cite{efde} to appear elsewhere.

\section{Conclusion and perspectives}
\label{sec5}

In this paper we shown an alternative method to calculate the background generated by cosmological DNS during their periodic or quasi-periodic phases. We used an analogy with a problem of Statistical Mechanics in order to perform such a calculation, as well as taking into account the temporal variation of the orbital parameters of the systems. In this method we can easily change the distribution functions and the parameters without the need of  modifying the formalism.

We adopted here a plane period distribution to study the influence of the time evolution of the orbital frequency on the spectrum of GWs generated by DNS.

In subsequent papers to appear elsewhere we will use the formalism developed here to: (a) calculate the background of GWs generated by black hole (BH) binaries and NS-BH binaries. Moreover, it will discussed, among other issues, how different SFRDs affect the spectrum of the background of GWs; (b) consider the GW spectra generated by compact systems in eccentric orbits; and (c) calculate the background of GWs generated by the coalescence of compact binary systems. In all these papers we will also discuss the detectability of the corresponding spectra by the present and forthcoming GW detectors.

\section*{Acknowledgments}
EFDE would like to thank Capes for support and JCNA would like to thank
FAPESP and CNPq for partial support. Last, but not least, we would like to 
thank the referee for his (her) useful suggestions and criticisms.

\end{document}